\documentclass[prl,
superscriptaddress,
twocolumn,
aps,
amsmath,amssymb,floatfix
]{revtex4-2}

\usepackage{sidecap}
\usepackage{placeins}
\usepackage{graphicx,svg}
\usepackage{dcolumn}
\usepackage{bm}
\usepackage{lipsum}
\usepackage{braket}
\usepackage[english]{babel}
\usepackage{amsmath}
\usepackage{sidecap}
\usepackage{xcolor}

\DeclareOldFontCommand{\bf}{\normalfont\bfseries}{\mathbf}
\DeclareOldFontCommand{\rm}{\normalfont\rmseries}{\mathrm}
\DeclareOldFontCommand{\tt}{\normalfont\ttseries}{\mathtt}

\begin{document}

\title{Elastic Wave Packets Crossing a Space-Time Interface}

\author{Alexandre Delory}
\email[Corresponding author: ]{alexandre.delory@espci.psl.eu}
\affiliation{Institut Langevin, ESPCI Paris, Université PSL, CNRS, 75005 Paris, France }%
\affiliation{Physique et M\'ecanique des Milieux H\'et\'erog\`enes, CNRS, ESPCI Paris, Universit\'e PSL, Sorbonne Universit\'e, Université de Paris Cit\'e, F-75005}%
\author{Claire Prada}
\affiliation{Institut Langevin, ESPCI Paris, Université PSL, CNRS, 75005 Paris, France }%
\author{Maxime Lanoy}
\affiliation{Laboratoire d'Acoustique de l'Université du Mans (LAUM), UMR 6613, Institut d'Acoustique - Graduate School (IA-GS), CNRS, Le Mans Université, 72085 Le Mans, France.}%
\author{Antonin Eddi}
\affiliation{Physique et M\'ecanique des Milieux H\'et\'erog\`enes, CNRS, ESPCI Paris, Universit\'e PSL, Sorbonne Universit\'e, Université de Paris Cit\'e, F-75005}%
\author{Mathias Fink}
\affiliation{Institut Langevin, ESPCI Paris, Université PSL, CNRS, 75005 Paris, France }%
\author{Fabrice Lemoult}
\affiliation{Institut Langevin, ESPCI Paris, Université PSL, CNRS, 75005 Paris, France }

\date{\today}

\begin{abstract}
The interaction between waves and evolving media challenges traditional conservation laws. We experimentally investigate the behavior of elastic wave packets crossing a moving interface that separates two media with distinct propagation properties, observing the non-invariance of wavelength and frequency. Our experimental setup employs an elastic strip whose local stretching can be dynamically altered by pulling one end at a constant velocity. By demonstrating that this dynamic configuration creates a spatio-temporal interface traveling along the strip, we confirm theoretical predictions regarding observed shifts when a wave packet crosses this interface. 
\end{abstract}

\maketitle


Wave propagation through moving interfaces is a fundamental subject in physics, with profound implications that extend far beyond laboratory experiments. 
A striking example of this phenomenon can be observed at a cosmic scale. In our expanding universe~\cite{hubble1929relation}, waves—whether light or gravitational—navigate regions of space-time whose properties are in constant flux~\cite{novello2004nonlinear,uzan2007acceleration}. This picture is intimately connected to the principles of general relativity~\cite{einstein1905elektrodynamik,norton2004einstein}, where the presence of a massive body can alter the curvature of space-time and lead to compelling effects such as gravitational lensing~\cite{refsdal1964gravitational,liebes1964gravitational}. Additionally, the temporal evolution of these massive objects leads to a dynamic environment which has been evidenced through the existence of gravitational waves~\cite{abbott2016observation}.

At human scale, several striking examples of waves interacting with a moving medium can also be reported. Probably the simplest one relies on an optical wave interacting with an acoustic one: the acousto-optic effect~\cite{chang1995acousto}. It finds applications for modulating light~\cite{clark1979acousto,dugan1997high} or for switching lasers~\cite{clarkson1991acousto}. 
At radiofrequencies, as early as 1958~\cite{cullen1958travelling}, it was demonstrated that periodically modulated inductances allow to amplify the wave in a transmission line.  
Additional examples of parametric amplification include antennas in Magnetic Resonance Imaging~\cite{syms2008three}, fiber optics for optical telecommunication~\cite{hansryd2002fiber}, or four wave mixing~\cite{yariv1977amplified,hellwarth1979theory}. The Faraday instability is also a striking example of parametric amplification of gravito-capillary waves~\cite{faraday1831forms}.
With the advent of metamaterials~\cite{simovski2020introduction,krushynska2023emerging} and more recently the reconfigurable metasurfaces~\cite{kaina2014shaping,liu2021reconfigurable}, introducing time-varying properties offers a new tool to design advanced devices~\cite{caloz2019spacetime,caloz2019spacetime2,engheta_2021}.

Despite the extensive theoretical explanations and studies of the underlying physics~\cite{fante1971transmission,dodonov1993quantum,milton2017field,huidobro_2019,caloz_2020,chen_2021,wen_2022}, these phenomena lack direct experimental evidence.  
In this article, 
we present an easy-to-implement experiment  that allows the observation of a unique event on a macroscopic scale: the crossing of a spatio-temporal interface by a wave. For this, we start with a soft elastic strip and suddenly displace one of its two ends. This creates a moving front of transition between a region at rest and an elongated one. Then, we study the passage of a flexural wave packet through this interface in two possible configurations: the wave and the interface propagate either in opposite directions or in the same direction. Our experimental results, which are supported with a theoretical model and demonstrate that both frequency and wavenumber are shifted by this interaction.

\paragraph*{Creating a space-time interface ---}
\begin{figure*}
    \centering
    \includegraphics[width=15.5cm]{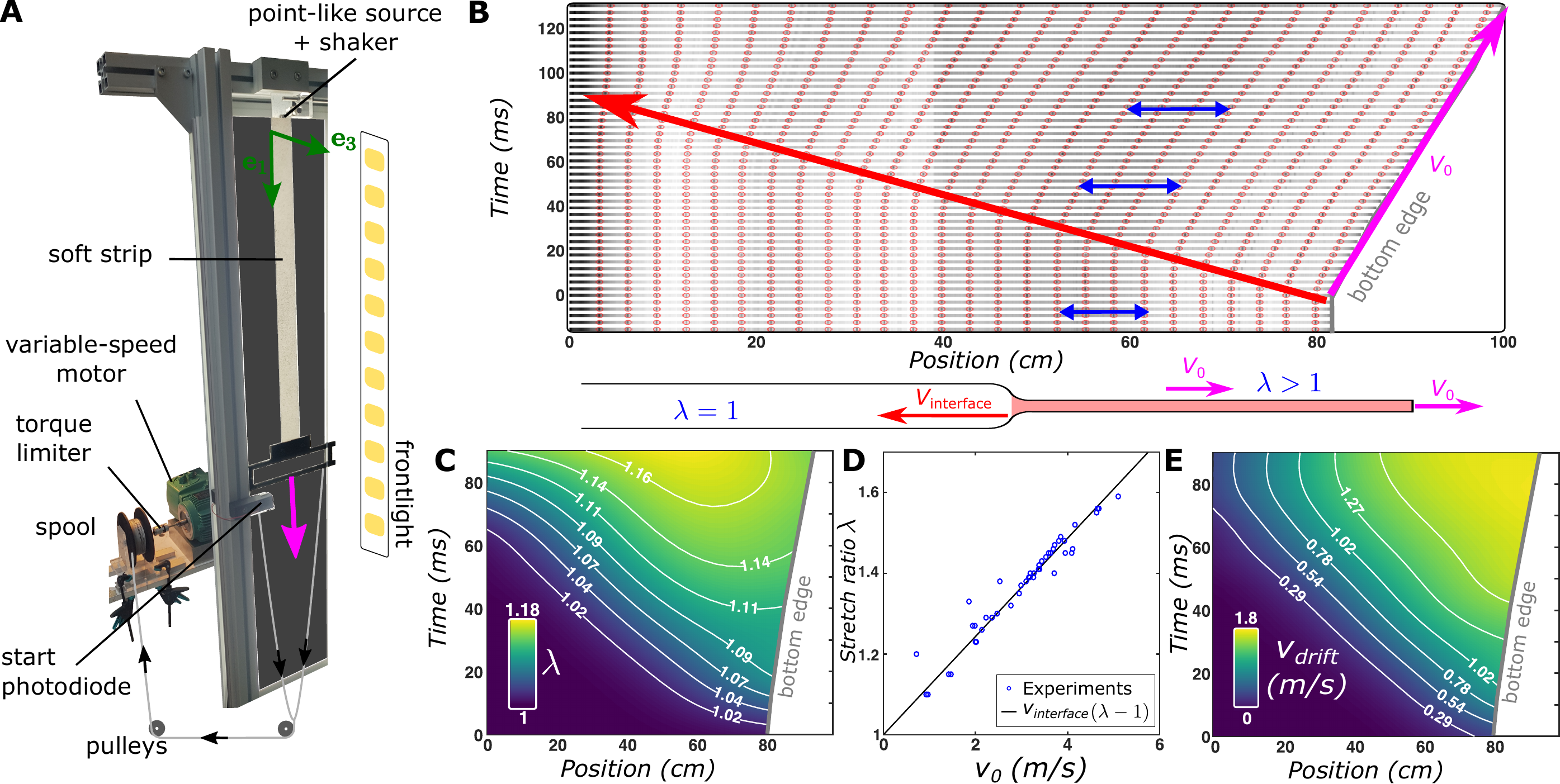}
    \caption{\textbf{Experimental setup and media characterization --- } 
    \textbf{A-} An elastic strip is attached at the upper end and pulled at a constant velocity $V_0$ at the lower end. The experiment is recorded with a camera at a frame rate of 1~kHz. The transverse displacement is extracted by tracking the position of the beam's edge, and the axial deformation by tracking the position in $x_1$ of darker spots on the edge. 
\textbf{B-} Spatio-temporal representation of the strip elongation. The bottom edge starts moving a $t=0$ with a speed $V_0$ (here 1.4 m/s), and an interface propagates leftward at speed $V_{\textrm{interface}}$ (as sketched below). The darker spots are shown as red circles and their position evolution are shown. The red arrow indicates the spatio-temporal interface moving at $V_{\textrm{interface}}$ here of approximately 8.5 m/s, with blue arrows highlighting the medium's stretching and drift relative to the laboratory frame. \textbf{C-} Map of the stretch ratio $\lambda$ extracted from the experiment B. \textbf{D-} Repeating the experiment for different $V_0$, a linear relation gives $V_{\textrm{interface}}=8.2$~m/s. \textbf{E-} Map of the particle velocity extracted from the experiment B.}
    \label{fig:setup}
\end{figure*}
The experimental platform relies on the propagation of a flexural wave along the direction $x_1$ in a rectangular beam of width $w$ and thickness $h$ in the plane $(x_2,x_3)$. 
This mode is ruled by the Euler-Bernoulli equation~\cite{doyle1989wave}:
\begin{equation}\label{eq:EulerBern}
   EI\frac{\partial^4u_3}{\partial {x_1}^4}-A\sigma\frac{\partial^2u_3}{\partial {x_1}^2}+A\rho \frac{\partial^2u_3}{\partial t^2}=0,
\end{equation}
\noindent with $u_3$ denoting the out-of plane displacement of the beam, $E$ the Young modulus of the material, $A=wh$ the beam's cross-section, $I=wh^3/12$ the second moment of area, $\sigma$ the applied longitudinal tensile stress and $\rho$ the mass density of the material. 
This equation immediately reveals that $\sigma$ acts as a tuning parameter: as $\sigma$ varies, a competition between the first and second terms explicitly emerges. At low stress, 
the propagation is governed by the flexural rigidity  $EI$. But, as the stress increases the tensile force $A\sigma$ begins to dominate.
Alternatively to the stress,  
this criterion can also be written in terms of the stretch ratio $\lambda$, invoking Hooke's law $\sigma = E(\lambda - 1)$. Note that this equation remains a simplified model compared to the full description of a true highly stretched beam~\cite{delory_sm_2024}. Notably, the hyperelastic behaviour as well as the rheology of the materials at play need to be considered for more quantitative measurements. Nevertheless, all these aspects could be incorporated in a Young modulus that would depend on both the frequency and the elongation ratio~\cite{delory_sm_2024}.

To create a space-time interface, it is necessary to modify this elongation while a wave packet propagates. 
As depicted in Figure~\ref{fig:setup}.A, a soft strip, made of Ecoflex-OO30  and of dimensions 80~cm~$\times$~1.1~mm~$\times$~3~cm, is clamped at its upper end, and connected to a variable-speed motor at its lower end via a rope. At time $t=0$, the motor is actuated and the bottom extremity starts stretching the strip. As can be seen from the spatio-temporal time lapse sequence of Figure~\ref{fig:setup}.B, the mobile edge then initiates a uniform motion which stretches the lower part of the beam at a constant velocity $V_0$ imposed by the motor. Interestingly, the rest of the strip remains totally still for a short delay that can reach up to 80~ms for the region located near the clamped end. The strain is not instantaneous across the entire strip: the longitudinal elongation travels with a given velocity which we now explicit.

\begin{figure*}
    \centering
    \includegraphics[width=16cm]{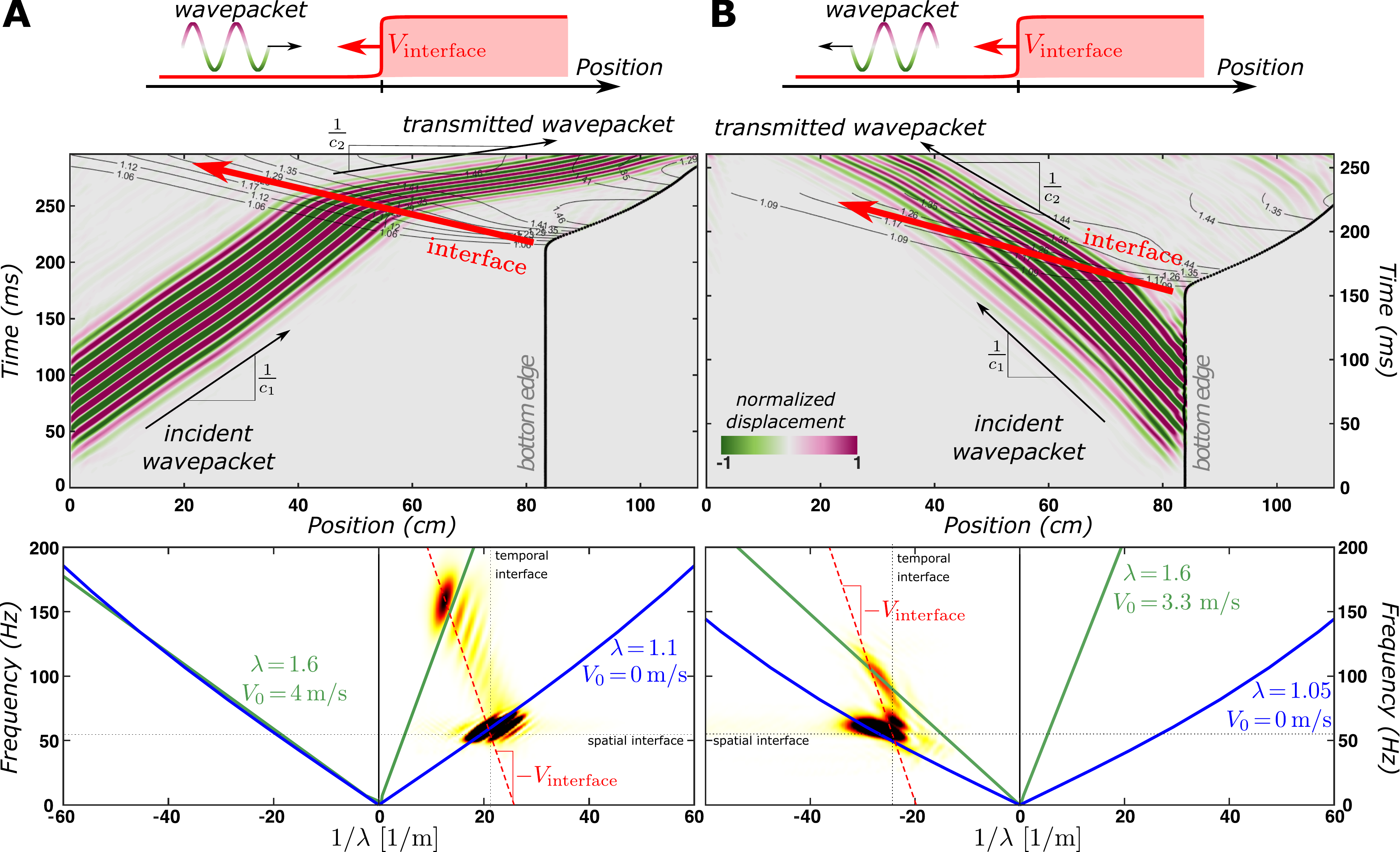}
    \caption{\textbf{Crossing a space-time interface ---} Space-time diagram (and reciprocal representation) of a flexural wave packet crossing a space-time interface propagating in the opposite direction (A) or in the same direction (B). The strip is pulled at 4 m/s in (A) and 3.3 m/s in (B). The non-invariance of both wavelength (or wavenumber) and period (or frequency) for the transmitted waves is demonstrated. Below are represented the spatio-temporal Fourier transforms of the same data. Superimposed are shown the theoretical dispersion curves before (blue) and after the interface (green).}
    \label{fig:impact}
\end{figure*}

This elongation wavefront, previously observed and studied in materials such as a rubber band~\cite{vermorel_2006} or a slinky~\cite{cross_2012}, corresponds to a longitudinal deformation of the strip. It is thus governed by a simpler wave equation~\cite{landau1986theory}:
\begin{equation}\label{eq:LongiWave}
    E\frac{\partial^2u_1}{\partial {x_1}^2}-\rho \frac{\partial^2u_1}{\partial t^2}=0,
\end{equation}
\noindent where $u_1$ now denotes the displacement along the direction $x_1$. 
The boundary condition in the experiment is simply expressed as $u_1(L,t)=V_0 t H(t)$, where $H(t)$  represents the unit step function. Through the use of Laplace transform, one can readily ascertain that the displacement $u_1$ writes:
\begin{equation}\label{eq:LongiDisplacement}
    u_1(x_1,t)=V_0\xi H\!(\xi) \quad \textrm{with} \quad \xi=t+\frac{x_1-L}{V_{\textrm{interface}}}.
\end{equation}
This displacement field indicates the propagation of an interface with a negative velocity. This corresponds to the so-called bar velocity~\cite{kynch1957fundamental} which is independent of the geometry of the beam: $V_{\textrm{interface}}=\sqrt{E/\rho}$.
After the interface has passed, the local elongation ratio takes the form:
\begin{equation}
\label{eq:LongiElongation}
  \lambda=1+\frac{\partial u_1}{\partial x_1}=1+\frac{V_0}{V_{\textrm{interface}}}.  
\end{equation} 
Figure~\ref{fig:setup}.C shows the strip's elongation map as a function of space and time. It demonstrates a spatiotemporal transition from initial rest to a stretched configuration. Because $\lambda$ explicitly depends on $V_0$, the contrast in elongation between the two regions can be controlled by tuning the motor velocity $V_0$. And, as discussed earlier, the amount of elongation directly sets the velocity of the in-plane flexural waves. In comparison to the theoretical model, the interface is not as sharp as a step function but rather exhibits a slower variation from 1 to 1.16. It can be attributed to the acceleration of the motor and to the rheology of the material~\cite{lanoy_PNAS_2020,delory_eml_2023,delory_sm_2024}.
A systematic extraction of the stretch ratio for various values of $V_0$ is presented in Figure~\ref{fig:setup}.D. Not only it confirms the possibility to create a wide variety of elongation ratios, but the agreement with equation~(\ref{eq:LongiElongation}) allows to retrieve a precise value of $V_{\textrm{interface}}=8.2$~m/s. This result is in good agreement with the expression of the bar velocity (see above), with $E=67$~kPa and $\rho=1000$~kg/m$^3$, as measured in a previous work~\cite{delory_sm_2024}.

A second crucial aspect to this experimental setup involves a peculiarity of the stretched medium. According to equation~(\ref{eq:LongiDisplacement}), after the interface passage, the medium experiences a uniform motion at velocity $V_0$. Consequently, not only the medium is stretched but it also undergoes a uniform translation. This appears readily on Figure~\ref{fig:setup}.B, where red dots and blue arrows indicate linear translation to the right. The experimental drifting velocity obtained from the experiment is color-coded in Figure~\ref{fig:setup}.E. 
After the interface passage, a homogeneous medium drifting at the constant velocity $V_0$ is observed in the upper right corner of the figure (once again the interface is not as sharp as a step function and a smooth transition is observed in the figure). This translation significantly affects the propagation of the transverse wave when observed in the laboratory frame. Specifically, due to the translation of the wave's support, a Doppler shift needs to be accounted for:
\begin{equation}
{c}_{\text{laboratory frame}} = c(\lambda) \pm V_{0}.
\end{equation}
This law breaks the spatial reciprocity~\cite{godin1997reciprocity,fleury2014sound,pelat2020acoustic}:  in the laboratory frame, a wave traveling in the $x_1$ direction will appear faster than
a wave that propagates in the negative direction. 

As a summary of this part, by pulling a strip at a constant velocity $V_0$, an interface is created. It propagates  from the bottom to the top of the strip at the velocity $V_{\textrm{interface}}=\sqrt{E/\rho}$  thus separating the strip between a region at rest and a region translated at constant velocity $V_0$ under a uniform strain.

\begin{SCfigure*}
\includegraphics[width=.75\textwidth]{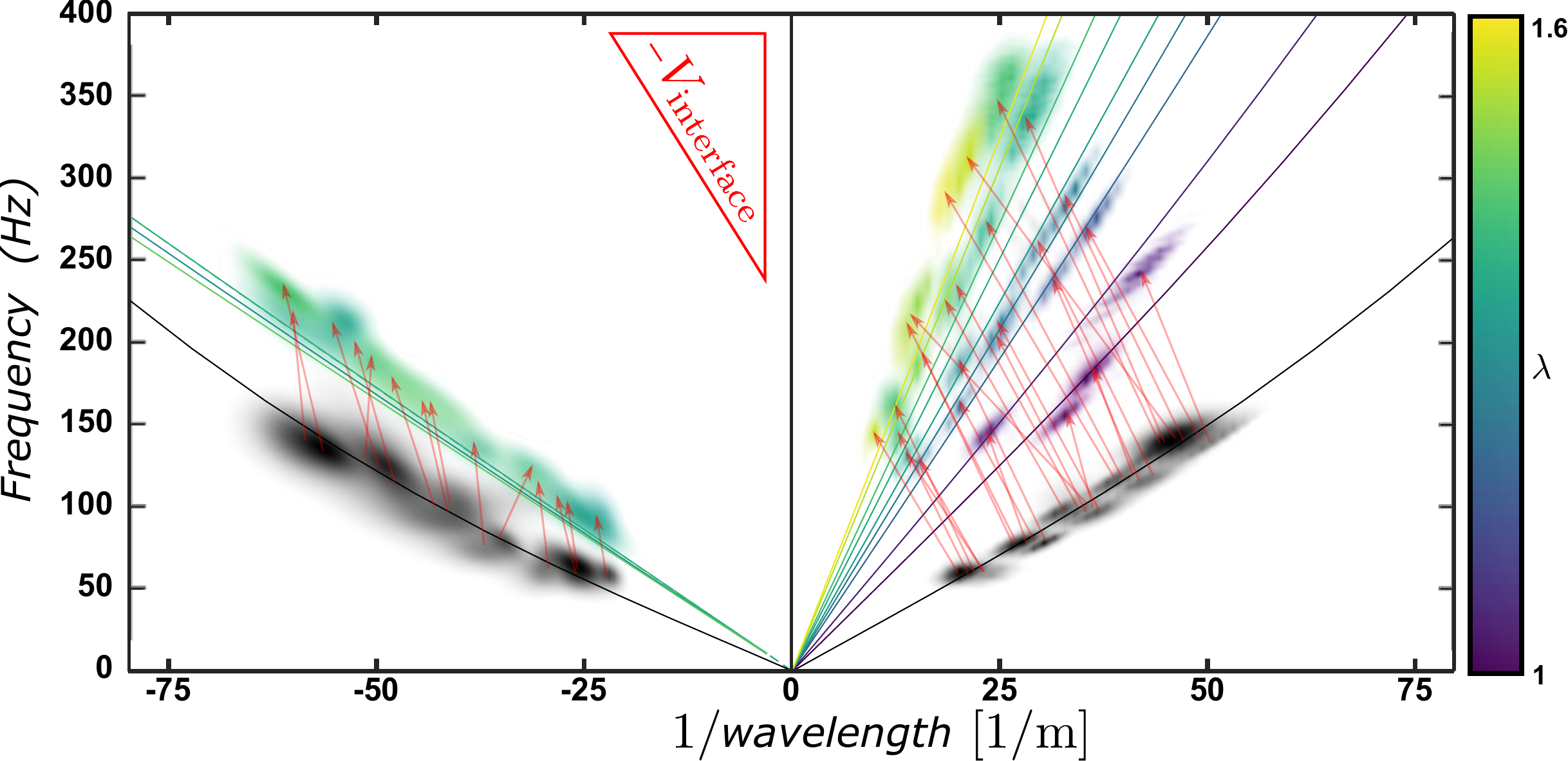}
  \caption{\textbf{Combined results of all experiments --- }  Frequency versus wavenumber representation, illustrating the alignment of frequency and wavenumber shifts with the slope $V_{\textrm{interface}}$ of equation~(\ref{eq:shifts}). Each red arrow corresponds to a given experiment, it points from the incident wave toward the transmitted wave. Lines indicate the theoretical predictions for various $\lambda$.\vspace{22pt}}\label{fig:expCross}
\end{SCfigure*}

\paragraph*{A wavepacket crossing the space-time interface ---}
We now study the interaction of a flexural wave with this interface.
A Tira Vib shaker pushes back and forth the strip in the $x_3$ direction thanks to a plastic holder. The camera records the strip's profile and allows to extract the space-time trajectory of the wavepacket (Figure~\ref{fig:impact}). Two different configurations are considered. For the first one depicted in Fig.~\ref{fig:impact}A, the wavepacket is generated in the area at rest and travels towards the stretched region. After 200~ms, the flexural wavepacket enters the expanded zone. Interestingly, the interface does not induce any reflection of the wavepacket. However, a transmitted wave is observed and it displays an increase in both its frequency, from 55~Hz to 168~Hz, and wavelength, from 4.5~cm to 8~cm. In the second configuration, the source is placed at the bottom and the flexural wave is generated before the stretching.  The wavepacket now travels in the same direction as the interface (Figure ~\ref{fig:impact}B). Starting from the resting region, it eventually gets overtaken by the interface that travels faster. Again, shifts in frequency and wavelength are observed: from 55~Hz to 100~Hz in frequency and from 4.5~cm to 3.8~cm in wavelength.

These results exemplify the paradigm shift introduced by considering a moving interface. Generally, the invariance of frequency (or equivalently, the temporal period) when crossing a spatial interface is justified by invoking the invariance by time translation along with the linearity of the medium. However, recent experiments involving a temporal interface~\cite{bacot2016time,apffel2022frequency}, echoing previous theoretical developments~\cite{mendoncca2002time,mendoncca2003temporal}, have revealed that the presence of a reflected time-reversed wave can be attributed to the conservation of the wavenumber (or wavelength). In our case, none of these quantities is preserved. 
It underscores the fundamental changes brought about by the moving interface~\cite{caloz_2020}. This phenomenon can be summarized by the fact that the change of wavenumber is linked to the change of pulsation with respect to the velocity of the interface. From our experimental observations, we evidence the following generalized conservation law:
\begin{equation}
\frac{\omega_2-\omega_1}{k_2-k_1}=-V_{\textrm{interface}},
\label{eq:shifts}
\end{equation}
\noindent where $\omega$ corresponds to the angular frequency and $k$ corresponds to the wavenumber of the wave, and is negative if the wave propagates in the negative direction. Note that this expression yields the usual conservation laws regarding space or time interfaces: $V_{\textrm{interface}}=0$ yields $\omega_2=\omega_1$ (space interface) and $V_{\textrm{interface}}=\infty$ yields $k_2=k_1$ (time interface).

Despite the broken time and space invariance in this system, the 2D Fourier Transform still yields valuable information about the spectral contents. The results are depicted below the spatio-temporal maps of the waves in Figure~\ref{fig:impact}. For each scenario, two peaks are observed in the reciprocal plane. For the wave propagating in the negative direction, the wavenumbers naturally correspond to  negative values. For the strip at rest (blue) as well as for the stretched and moving one (green), all the peaks coincide with the full 3D calculation of the dispersion relations~\cite{delory_sm_2024,kieferGEW}. These calculations fully incorporate the acoustoelastic effect and the non-Hookean nature of the strip. They are more accurate than the oversimplification made in equation~(\ref{eq:EulerBern}) and permit to describe accurately the dispersion relation of Medium 2. The transition in the dispersion diagram from the curve of Medium 1 
toward the curve of Medium 2 follows the line of slope $-V_{\textrm{interface}}$ and confirms the relation between the wavenumber and frequency shifts of equation~(\ref{eq:shifts}).

Eventually, the experiment is repeated for various pulling velocities and wavepacket's frequencies. For each experiment 2D space-time Fourier transforms are applied and the relevant spots of the Fourier transforms are kept. All the results are accumulated and are summarized on the same axes in Figure~\ref{fig:expCross}. The initial wavepackets are all aligned with the dispersion relation of flexural waves in a strip almost at rest (black line). A slight difference between the negative and positive wavenumbers is visible because of the gravity that exerts a smaller elongation of the strip at its bottom compared to the top.  After crossing the interface, all wavepackets show a change in frequency and wavenumbers. The transmitted wave ends on the dispersion curve of the Medium 2 which depends on the pulling velocity $V_0$ with a good agreement with the theoretical predictions (lines on the figure). The transition between the incident wave packet and the transmitted one follows the equation~(\ref{eq:shifts}) as indicated by the red arrows that are all parallel with the same slope of $-V_{\textrm{interface}}$. 

\paragraph*{Discussion ---}
This simple macroscopic experiment effectively demonstrates the key feature of a spatio-temporal interface: the non-conservation of wavenumber and frequency. However, some comments regarding this platform are needed. 

Firstly, the experiment did not reveal a reflective wave. This might suggest that the interface is not a sharp interface but a smooth adiabatic one. 
Or alternatively, we can presume that it is a consequence of the matched impedances of the two media. By considering the transverse wave is similar to that in a vibrating string, we can assume the impedance to be $\sqrt{\sigma\rho}$~\cite{BerkeleyWaves}. In Medium 2, since it is stretched and the material is nearly incompressible, the linear density is divided by $\lambda$, while the stress is increased by a factor of $\lambda$ according to Hooke's law. Moving forward, we aim to adjust the platform to allow impedance mismatch and thus favour the observation of reflections.

Secondly, in this experiment the wave is convected by the motion of the support, resulting in a substantial frequency shift 
observed in the scenario of figure~\ref{fig:impact}A where the wavepacket travels in the opposite direction compared to the interface. This property can be found in other wave phenomena, such as gravity-capillary waves at the surface of moving fluids~\cite{shyu1990blockage,rousseaux2008observation}, acoustic waves in flowing pipes~\cite{godin1997reciprocity,pelat2020acoustic}, or elastic waves in waveguides coupled to flowing fluids like arteries~\cite{baranger2023fundamental}. These examples could also be promising candidates for investigating static boundaries in the laboratory frame, which would be in translation in the moving frame.

Lastly, perhaps the most compelling example of a time-evolving medium is the expanding universe. Consequently, we should consider higher dimensions for observing refraction at the interface. Transitioning from a strip to a membrane, and thereby moving to a two-dimensional medium, is relatively straightforward and will be the focus of future investigations. However, the challenge remains in 3D, given the contrast between longitudinal and transverse wave velocities. The solution may lie in engineering materials within the framework of bistable materials~\cite{nadkarni2016unidirectional,zareei2020harnessing}.

\paragraph*{Conclusion --- }
This Letter presents an experiment where a wave packet crosses a moving interface separating two media with distinct propagation properties. This experiment relies on the propagation of a transverse wave in an elastic strip which is dynamically stretched. The observed transmitted wave in our setup exhibits both a frequency shift and a wavenumber shift, driven by the velocity of the interface and the propagation characteristics of the two media.

While our experiment has yielded valuable insights, there are areas for further investigation, particularly concerning the amplitudes' coefficients and impedance considerations. Also, the step function is only the elementary brick of more complex transient functions. Introducing a time dependency in the pulling velocity $V_0(t)$ should permit to investigate complex space-time variations as well as connecting to parametric amplification experiments~\cite{faraday1831forms,douady1990experimental}. Our ongoing objective is to refine the experiment to deepen our understanding of these unusual wave phenomena, especially regarding the non-conservation of energy. 

\paragraph*{Acknowledgments --- }This work has received support under the program ``Investissements d’Avenir'' launched by the French Government and partially by the Simons Foundation/Collaboration on Symmetry-Driven Extreme Wave Phenomena.  A. D. acknowledges funding from French Direction Générale de l'Armement.


%

\end{document}